# Layered Depth-Normal Images: a Sparse Implicit Representation of Solid Models

Charlie C. L. Wang[*]

Department of Mechanical and Automation Engineering
The Chinese University of Hong Kong

Yong Chen

Department of Industrial and Systems Engineering
University of Southern California

**Abstract**
This paper presents a novel implicit representation of solid models. With this representation, every solid model can be effectively presented by three layered depth-normal images (LDNIs) that are perpendicular to three orthogonal axes respectively. The layered depth-normal images for a solid model, whose boundary is presented by a polygonal mesh, can be generated efficiently with help of the graphics hardware accelerated sampling. Based on this implicit representation – LDNIs, solid modeling operations including the Boolean operations and the offsetting operation have been developed. A contouring algorithm is also introduced in this paper to generate thin structure and sharp feature preserved mesh surfaces from the layered depth-normal images. Comparisons between LDNIs and other implicit representation of solid models are given at the end of the paper to demonstrate the advantages of LDNIs.

## 1. Introduction

Geometric modeling on solid models in implicit representation (or volumetric representation) has been widely investigated in the applications of computer graphics [1-3]. Benefited from the compact and intuitive mathematical representation, the solid modeling operations developed for implicit representations are usually robust and easy to implement. However, although the implicit representation on uniform grids (e.g., distance-field [1]) can be generated by the ray casting method and can be even further speed up with the help of graphics hardware, it is a dense representation which is memory consuming. The memory management of algorithms generating adaptive grids is more efficient, but usually much more complex and needs more computing time. These two factors restrict the employment of implicit representation in CAD/CAM applications.

---

[*] Corresponding Author. E-mail: cwang@mae.cuhk.edu.hk; Fax: (852) 2603 6002.

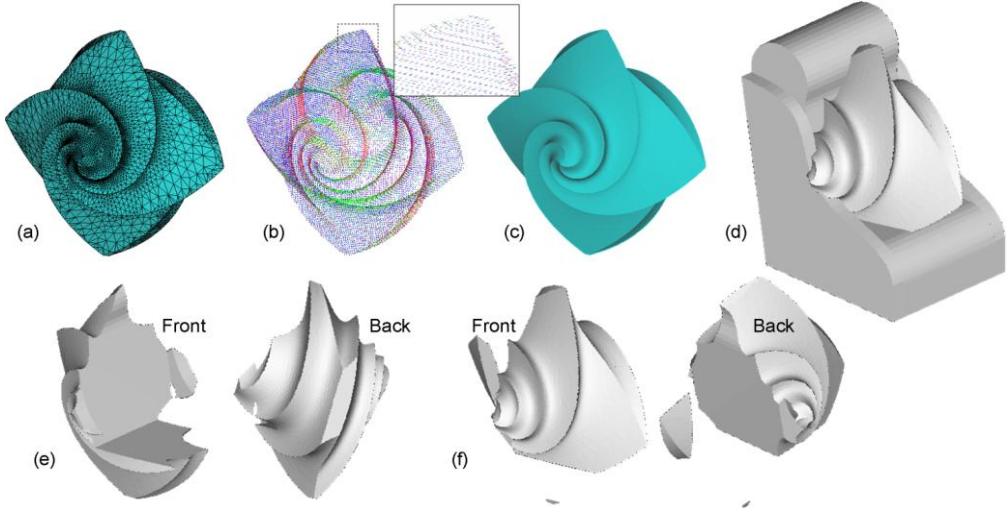

**Figure 1**: An example of solid modeling based on layer depth-normal images (LDNIs): (a) the given octa-flower model in polygonal mesh, (b) example LDNIs of the octa-flower model sampling at the rate of 128×128 where red, blue and green colors present samples on the LDNIs perpendicular to $x$-, $y$- and $z$-axis respectively (computed in 0.141 sec), (c) the contouring result from the 128×128 LDNIs (computed in 2.703 sec), (d) the result of Boolean operation on 512×512 LDNIs – octa-flower $\bigcup$ chair (computed in 2.267 sec), (e) the result of Boolean operation on 512×512 LDNIs – octa-flower $\bigcap$ chair (computed in 1.279 sec), and (f) the result of Boolean operation on 512×512 LDNIs – octa-flower \ chair (computed in 1.078 sec). Note that (d)-(f) are rendered directly from LDNIs by point-based rendering techniques [26].

Based on the above reasons, we are going to develop a new implicit representation – Layered Depth-Normal Images (LDNIs), which can achieve a balance of required memory and computing time. By this representation, every solid model can be effectively represented by three layered depth-normal images (LDNIs) that are perpendicular to three orthogonal axes respectively. For a given sampling rate $w$, the required memory of LDNIs is only $O(w^2)$ for most practical models which is similar to the adaptively sampled implicit representations but more compact. Besides, the construction of LDNIs from closed 2-manifold polygonal meshes can be efficiently completed by a rasterization technique implemented with the help of graphics hardware. Solid modeling operations on LDNIs have also been developed. Solid models in many CAD/CAM applications still need to have the boundary representation (B-rep) because many current available CAD/CAM techniques (e.g., CNC tool path generation, rapid prototyping, parting line generation of mold design, etc.) are based on B-rep. Many techniques, such as Marching Cubes algorithms [4, 1, 5-8] and Dual contouring algorithms [3, 9-12], have been developed to convert an implicitly represented model into a polygonal mesh model (i.e., B-rep). We develop a contouring algorithm for LDNIs akin to dual contouring [3, 9] but with necessary modifications to preserve both sharp features and 2-manifold mesh on resultant surfaces. Note that the contouring algorithm is only applied on LDNIs when the downstream applications need the boundary representation as input. For the repeating solid modeling operations, e.g., union operations in the computation of Minkowski sum or sweeping, only LDNI-representation is needed. The temporary results on LDNI representation can be directly rendered by point-based rendering techniques [26]. Figure 1 shows an example for the Boolean operations on LDNIs and the direct rendering of LDNIs.

### 1.1. Related work

Solid modeling based on B-rep has been investigated for many years. Most of the existing approaches are based on the intersection calculation followed by a direct manipulation of boundary representation. Surveys can be found in [13] and [14]. The topology correctness of resultant models relies on the robust intersection computation (ref. [15]). Although some recent published approach [16] proposed a topologically robust algorithm for Boolean operations using approximate arithmetic, the computation still needs to face the topology regularization problem in some extreme cases (e.g., the examples shown in [17]). Therefore, many approaches [18-20] adopted volumetric representation to approximate the operations of solid modeling since the computation of solid modeling operations (such as, Boolean operations and offsetting operation) on volumetric representation is more robust, compact and easy to implement. The solid modeling framework introduced in this paper is also based on a volumetric representation; however, the representation of LDNIs is sparser (like a sparse matrix) so that it is efficient in memory cost.

The simplest volumetric representation of solid models is voxel-based (ref. [21]). However, as the binary voxels cannot give a good representation of smooth surface and sharp features, the methods based on distance-fields are soon employed to replace binary voxels. A survey of 3D distance-field techniques can be found in [22]. However, sharp edges and corners are removed during the sampling of uniform distance-fields. Over-sampling could somewhat reduce the aliasing error by taking the cost of increasing storage memory in uniform sampling or by taking the cost of more computing time in adaptive sampling. Furthermore, as being observed by Kobbelt et al. in [1], even if an over-sampling is applied, the associated aliasing error will not be absolutely eliminated since the surface normals in the reconstructed model usually do not converge to the normal field of the original model. Based on this reason, recently developed volumetric approaches always encode both the distance from a grid node to the surface under sampling and the normal vector at the nearest surface point to the grid node (see [9, 10, 12, 19]) – which is called Hermite data. The layered depth-normal images (LDNIs) proposed in this paper also encode Hermite data points during the sampling procedure. Like [1, 3], we do not encode Hermite data on grid nodes but on surface intersection points of ray casting. We develop a method based on [23] and [24] to accelerate the encoding of Hermite data on LDNIs by the graphics hardware.

Many CAD/CAM applications adopt B-rep to describe the shape of a solid model in their systems. Thus, it is very important to have an algorithm to generate 2-manifold polygonal mesh surface (B-rep) from a volumetrically represented solid model. Marching Cubes algorithm (MC) [4] is the most widely used approach in literature to generate triangular mesh surface from an implicit surface. The original MC algorithm [13] may produce isosurfaces with holes due to topologically inconsistent decisions on the reconstruction of ambiguous faces, where the borders used by one incident cube do not match the borders of the other incident cube. Many methods have been developed for solving this problem (e.g., [5, 8, 11, 25]). Another drawback of MC and its variants is that the resultant mesh surfaces always lack sharp edges and corners. Therefore, some approaches [3, 9-12, 19] employed the dual contouring techniques to generate an isosurface on Hermite data. Based on the basic idea of dual contouring in [3], the authors of [9] and [10] developed an extension of dual contouring that can reconstruct thin-structures that will be missed by the original dual contouring algorithm. The same authors of [3] recently developed a new extension of dual contouring in [12] that can ensure the two-manifold output of isosurfaces, which is a necessary property to CAD/CAM applications. They conduct the MC table of Nielson in [11] to guarantee the two-manifold property. However, when using the MC table of Nielson in [11], it is impossible to reconstruct the shells whose thickness is less than the width of grids, which is an intolerable drawback of CAD/CAM applications. Moreover, recently in [40], a non-manifold case is found on the MC table of Nielson. Here, we develop a LDNI-to-mesh conversion algorithm derived from the dual contouring that can capture both sharp edges/corners and shin structures. Our LDNI-to-mesh conversion algorithm is equipped with two-manifold correction techniques.

The purpose of techniques presented in this paper is different from the point-sampled geometry approaches [26-28], which focused on the interactive rendering. In these approaches, the shape of a model is described by a set of surface points coupled with normals (i.e., surfel). As aforementioned, the CAD/CAM applications such as CNC and rapid prototyping planning need to have B-rep of solid models. Although we can generate B-rep from the surfels, the structural information of samples, that can be used to speed up the solid modeling operations and the contouring process, is missed. Moreover, the point-sample geometry does not give an efficient way to evaluate the *inside/outside* of a point. Similar to point-sampled geometry, the surface information of solids encoded by our LDNIs is also stored by a set of points coupled with normal vectors. However, the points in LDNIs are well organized in a data structure so that the following solid modeling operations and contouring can be implemented easily and completed in an interactive speed. Recently, Nielson and Museth [29] developed an efficient data structure – DT-Grid for representing high resolution level sets. It is not clear whether it can be easily applied in solid modeling. Their method cannot preserve sharp feature as no Hermite data is recorded, and the conversion algorithms between mesh and DT-Grid representations have not been given.

With the development of graphics hardware, more and more computer graphics applications now employ GPU to speed up the computation of bottle-neck steps (e.g., rendering, collision detection, voxelization, and distance-field computation). The group of Manocha developed several GPU-based approaches for the distance-field evaluation [30-32]. However, the approaches cannot be directly used in the implicit representation based solid modeling as the uniformly sampled distance-field can hardly capture the sharp features and thin-shells. The authors in [24] also developed the collision detection approach for solid models based on the decomposition of Layered Depth Images (LDI) [33] that can be accelerated in graphics hardware by using OpenGL. LDI was originally introduced as an efficient image-based rendering technique. An LDI can be used to approximate the volume of an object since the LDI data structure essentially store multiple depth values per pixel. Stimulated by their approach, we have developed the sparse implicit representation, Layered Depth-Normal Images (LDNIs), for solid models in this paper. The technique developed here is different from [34] and [35], in which the major purpose is to render CSG tree of primitives instead of computing the resultant shape of solid modeling operations on models.

The approach proposed in this paper is somewhat similar to the Ray-rep in the solid modeling literature [36-38]. Menon and Voelcker sampled the solid models into parallel rays tagged with h-tag (i.e., the information of half-space at the endpoints of rays) in [37] so that the completeness of Ray-rep can be generated. The conversion algorithm between Ray-rep and B-rep or CSG is also given in [37]. As mentioned in [38], Ray-rep can make problem easy in the applications involving offsets, sweeps, and Minkowski operations. However, different from our LDNIs, the Ray-rep only stores depth values without surface normals in one ray direction. Furthermore, the algorithm presented in [37] to convert models from Ray-rep to B-rep does not take the advantage of structurally stored information so that it involves a lot of global search and could be very time-consuming.

Another line of research related to our work is the so-called Marching Intersections (MI) approach [39-41]. The representation of MI is similar to our LDNI representation but MI does not use normal vectors at samples in the surface reconstruction. This leads to the first deficiency of MI. As discussed in [1], aliasing error cannot be eliminated along sharp edges without the normal vectors (e.g., the sharp edges are chamfered in Figure 15(b)). The second deficiency of MI is that their surface reconstruction algorithm misses the features whose size is less than the width of a cell. However, our contouring method will overcome this by using complex-edges in our flooding based node clustering in each cell (e.g., the thin-structure in Figure 5(b)-(d) and 11). Moreover, the speed up by graphics hardware has not been addressed in their approaches.

### 1.2. Contributions

This paper has the following technical contributions:

- We introduce a novel sampling based sparse implicit representation of 3D solid models, Layered Depth-Normal Images (LDNIs), whose required memory is $O(w^2)$ with the sampling resolution $w$ for most practical models. The construction of LDNIs can be completed in an interactive speed with the help of graphics hardware, and the information encoded can be employed to reconstruct sharp edges/corners and thin-shells.
- General solid modeling operations, such as Boolean operations and offsetting operations, have also been developed for LDNIs. As the Hermite data stored in LDNIs is well structured, all operations can be easily implemented in parallel processes and/or on graphics hardware.
- Based on existing techniques, we have developed a modified contouring algorithm that can reconstruct thin sharp features. Two-manifold correction technique has been developed to fix non-manifold entities.

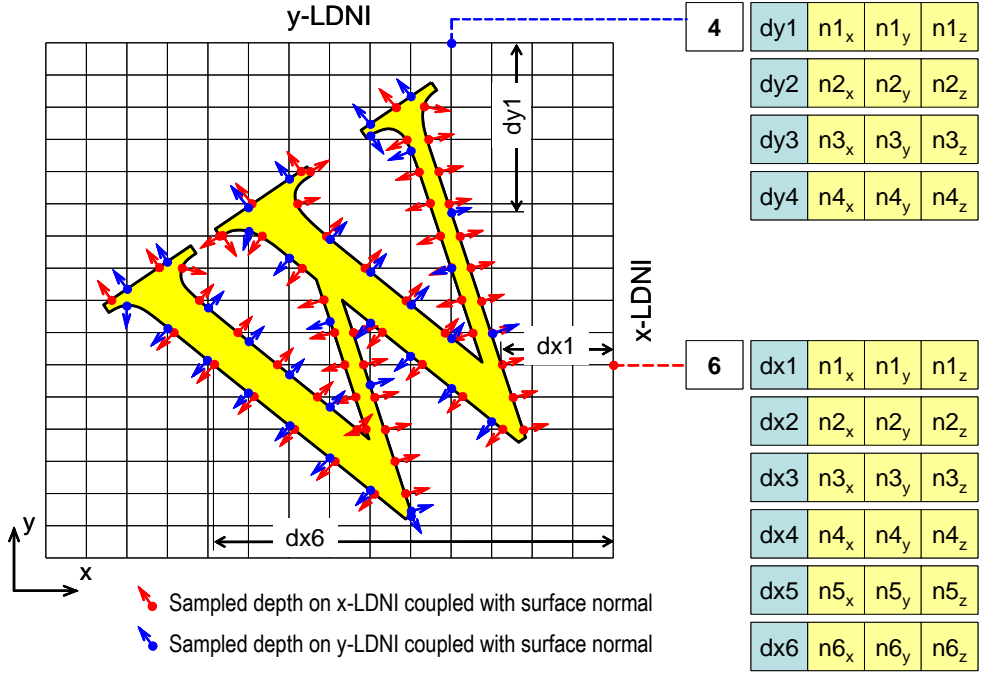

**Figure 2**: A two-dimensional illustration of Layered Depth-Normal Images (LDNI), where the dots represents the location of sampled depth and the arrow denotes the unit surface normal vector at this point. Red color is employed for the x-LDNI that is perpendicular to *x*-axis, and blue is for y-LDNI.

## 2. Layered Depth-Normal Images (LDNIs)

We propose a new representation, called Layered Depth-Normal Images (LDNIs), to implicitly encode the shape of solid models as a structured collection of Hermite data.

**Definition 1**   A single Layered Depth Image (LDI) with a specified viewing direction is a two-dimensional image with $w \times w$ pixels, where each pixel contains a sequence of numbers that specify the depths from the intersections (between a ray passing through the center of pixel along the viewing direction and the surface to be sampled) to the viewing plane and the depths are sorted in the ascending order.

Note that the intersections here exclude the case that a ray is parallel to the intersected faces.

**Definition 2**   A single Layered Depth-Normal Image (LDNI) is an extension of LDI where each depth is coupled with the unit normal vector of the sampled surface at the intersection point: x-LDNI is a Layered Depth-Normal Image viewed along the inversed direction of *x*-axis (i.e., the LDNI is perpendicular to *x*-axis), and y-LDNI and z-LDNI are perpendicular to *y*- and *z*-axis respectively.

**Remark 1**   An edge is defined as *silhouette-edge* if only one of its adjacent polygons faces along the current viewing direction.

When a ray intersects an edge shared by two faces, no intersection will be counted if this edge is a *silhouette-edge* and one intersection will be sampled for the *non-silhouette-edges*. For a *non-* *silhouette-edge*, the normal vector at either of its two adjacent faces will be selected and encoded randomly.

**Definition 3**   A structured set of Layered Depth-Normal Images (LDNIs) consists of x-LDNI, y-LDNI and z-LDNI with the same resolution $w \times w$, and the images are located to let the intersections of their rays intersect at the $w \times w \times w$ nodes of uniform grids in $\Re^3$.

Figure 2 gives a two-dimensional illustration of LDNIs, where the red dots and arrows indicates the Hermite data points recorded on the x-LDNI and the blue ones illustrate the Hermite data points on the y-LDNI. The example information stored in one pixel on the x-LDNI (linked by the red dash line) and one pixel on the y-LDNI (linked by the blue dash line) is also illustrated in Figure 2, in which the slots with blue background present the depth values and the yellow slots denotes unit normal vectors. From Definition 1-3, we can find that the information stored in LDNI is different from other uniform sampled implicit representation – here only the set of Hermite data points on the surface of a model are sparsely sampled and stored. The stored sample data likes the elements in sparse matrices. That's why we consider LDNIs as a sparse implicit representation.

**Remark 2**   The boundary surface of a solid model will not self-intersect.

**Definition 4**   For a correctly sampled solid model represented by Layered Depth-Normal Images, the number of sampled depths on a pixel must be even.

Note that when using graphics hardware accelerated method to obtain LDNIs, the guarantee of this property is based on the implementation of rasterization on the hardware. According to our experimental tests, even number of intersections is always reported when the mesh surface of input solid modes is closed. Moreover, the self-intersections on closed mesh surfaces can be eliminated by the method in [42].

**Proposition 1**   For the Layered Depth-Normal Images with pixel width *d*, a gap or thin-shell on the solid model whose thickness is less than *d* may be missed on images that are perpendicular to the gap or the thin-shell.

Proof of the above proposition is straightforward. By remark 2 and definition 4, we know that the range between the odd depth and the next even depth in one pixel of LDNI presents the volume inside the solid model under sampling. From proposition 1, it is easy to understand the reason why we need to have three orthogonal LDNIs to record the solid models with thin features.

The information stored in a pixel with the size ranges from $O(1)$ to $O(k)$ where *k* is the maximal number of layers of the model from this viewing direction. On most practical models, *k* is a constant number that satisfies $k << w$; in the worst case, $k \rightarrow w$ on all pixels, the upper bound of LDNI's memory complexity, $O(w^3)$, is reached. Therefore, we have the following proposition.

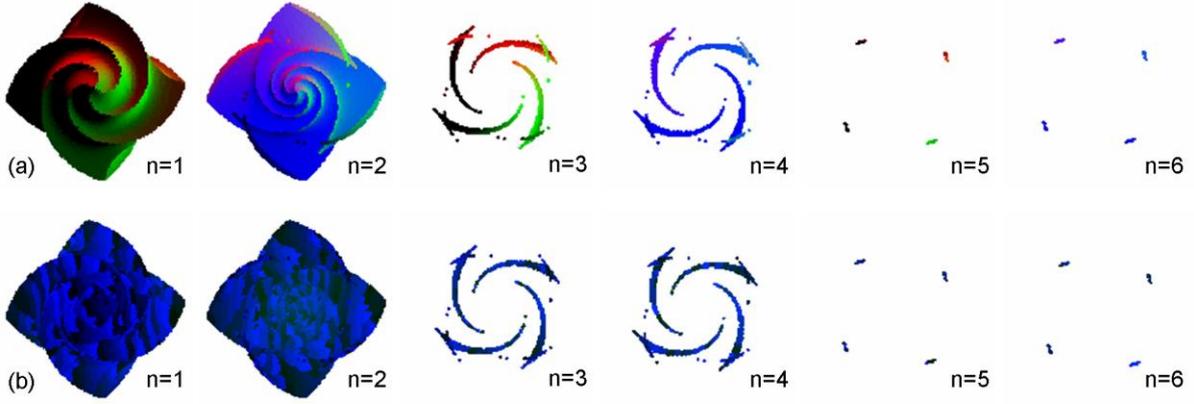

**Figure 3**: The layered images (z-LDNI with resolution: 128×128) generated by the multiple-pass rendering algorithm for the octa-flower model: (a) the images from layer 1 to layer 6 with surface normal encoded in RGB – approximate normal sampling, and (b) the images with face indices encoded as RGB colors – accurate normal sampling.

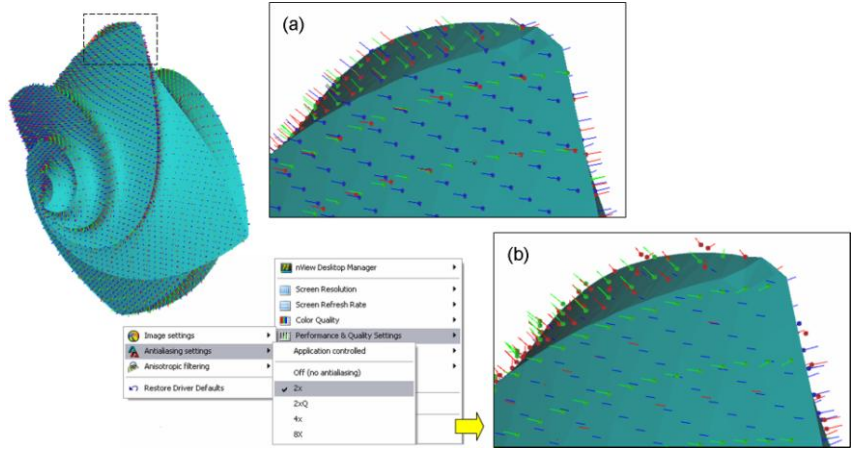

**Figure 4**: The anti-aliasing function provided by graphics hardware will make the sampling inaccurate. (a) The sampling result by turning off the anti-aliasing function – all samples are located on the surface of the given model. (b) By turning on the anti-aliasing function provided in the driver of graphics hardware but still keeping the anti-aliasing function in OpenGL off, the samples become away from the surface – some are shifted outside and some are inside. The test is evaluated on a NVIDIA GeForce 7600 GS graphics card.

**Proposition 2** The memory complexity of LDNI is $O(w^2)$ on most practical models, and with $O(w^3)$ in the worst case.

**3. Construction of LDNIs from Polygonal Mesh Using Graphics Accelerated Hardware**

Using graphics accelerated hardware to construct LDNIs for a solid model $H$ is similar to the well-known scan-conversion algorithm that the scan line along view direction alternately passes $H$ between interior and exterior. Similar to the sampling of LDI in [24], in order to generate a LDNI with the help of graphics hardware, the boundary surface mesh of $H$ has to be rendered multiple times.

The viewing parameters are determined by the working envelope, which is slightly larger than the bounding box of the model. Orthogonal projection is adopted for rendering so that the intersection points from parallel rays can be generated.

The repeated times of rendering are determined by the depth complexity $n_{max}$ of the model $H$ with the given direction (e.g., the model in Figure 2 is with $n_{max} = 8$ and $n_{max} = 6$ for x-LDNI and y-LDNI respectively). The value of depth complexity $n_p$ at every pixel can be read from the stencil buffer after the first rendering, in which the stencil test configuration allows only the first fragment to pass per pixel but still increment the stencil buffer in the later fragment pass. After that, $n_{max} = \max(n_p)$ can be determined by searching $n_p$ on all pixels and the depth values of the first pass fragments are stored in the depth buffer. If $n_{max} > 1$, additional rendering passes $n = 2$ to $n_{max}$ will generate the remaining layers and the stencil test configuration allows only the $n$-th fragment to pass. For the pixels with $n_p < n_{max}$, layers from $(n_p + 1)$ to $n_{max}$ do not contain valid depth values and are neglected. The depth values in the depth buffer are floating-point for most graphics cards. The above algorithm generates an unsorted LDI as fragments are in general rendered in arbitrary. Therefore, a post-sorting step is needed for the depths at each pixel. In order to avoid repeatedly sending the geometry and connectivity data from the main memory to the graphics hardware during the sampling – such data communication takes a lot of time for complex models, we compile a glList onto the graphics card and call the list for rendering the models repeatedly (ref. [43]). By this way, we only send the model to be sampled through the data communication bottleneck once, which greatly speeds up the sampling procedure.

The above algorithm for LDI does not record the surface normal at each sample. To solve this problem, we choose one of the following methods to encode the surface normal vectors during the decomposition of layered images.

1) **Approximate Surface Normal Sampling:** Three components of the unit surface normal vector at every polygon is mapped into three colors in RGB and rendered. Thus, the values of surface normal at each sampled fragment can be read back from the color buffer. However, the implementation of graphics hardware commonly gives only one byte for each component per pixel in the color buffer, so that the accuracy of this sampling method is very limited (i.e., the resolution is only 256 for the range [-1, 1]).

2) **Accurate Surface Normal Sampling:** In order to have accurate surface normal vectors encoded on each fragment, we first assign a unique ID to every polygonal face. The number of ID for each face is then mapped into a RGB-color. Therefore, after rendering all faces by the colors according to their IDs, we can easily identify which face contributes to a sample fragment by the RGB-color so that the surface normal on the face can be retrieved from the input model and then encoded at the sample. As each color component is with 8 bits, we can render up to $2^{24}$ distinguishable triangles into the frame buffer, which is usually much more than the required number of practical use.

Pseudo-code for the implementation of this algorithm in OpenGL is given in Appendix A. Figure 3 shows the layered images generated by these two methods for the octa-flower model. We can find that the images for the last few layers become more and more sparse – this is because that most pixels are with $n_p < n_{max}$.

When computing the coordinate in $\Re^3$ inversely from the sampled LDNIs, the only difficulty is to assign real coordinate values to integer indexed pixel location. Stimulated by [23], we compute the real coordinate of a pixel by its center instead of its lower left corners. This is because that the graphics hardware computes the scan-conversion from the center of pixels. Moreover, the graphics hardware accelerated scan-conversion is usually affected by the anti-aliasing function. First of all, all application controlled anti-aliasing functions should be turned off before the sampling. However, the driver of some hardware provides automatic anti-aliasing without considering the software configuration. Such anti-aliasing will introduce errors to the sampling (e.g., the one given in Figure 4(b)). Therefore, both anti-aliasing functions in OpenGL and the driver of graphics card should be turned off before sampling.

**Correctness of LDNIs sampled with the help of graphics hardware**

From experimental test, we find that the sampled results by using above hardware accelerated method on modern graphics cards satisfy the correctness property of solids represented in LDNIs (i.e., Definition 4). More specifically, when a ray intersects the non-silhouette-edge, only one intersection is counted in the stencil buffer. The value in the stencil buffer will be increased by two if the ray passing the center of a pixel intersects a silhouette edge. Based on the assumption in remark 2, the sorted Hermite samples at each pixel in LDNIs correctly represent the sampled solid along the ray passing this pixel's center. Samples from self-intersected solids can be correct by the self-intersection elimination method presented in [42] as long as the boundary of input solid is closed. In short, if the input model is valid, the sampling result is consistent with what we expect in the LDNI-based representation (defined in the previous section). The only difference between the theoretical representation and the sampling result from OpenGL based rasterization is the accuracy at samples which will be discussed below.

**Balance between time and accuracy**

In the *Approximate Surface Normal Sampling* method, 8 bits are used to encode each component of the normal vectors at a sample. The modern graphics cards start to support 32 bits for a color component, which can of course give more accurate sampling results. However, in our current implementation, the pixel information at each layer of the LDNIs needs to send back to the main memory to be further processed. Thus, using 32 bits for each color channel will slow down the sampling process. Furthermore, as will be shown later in the experimental result section, there will not too much difference in terms of shape approximation error when using 8 bits based normal vectors versus the accurate normal vectors calculated from the input mesh surface (i.e., by the *Accurate Surface Normal Sampling* method). Although the input normal vectors at vertices are normalized, the normal vectors generated by interpolation on a polygonal face will not be unit vector. Therefore, we read back all three components from graphics hardware instead of two. 32 bits based depth buffer is adopted during sampling (i.e., the float-point number) and transferred back into the main memory. Moreover, our experimental tests find that the improvement in terms of accuracy made by increasing the number of samples is more than the improvement made by increasing the precision of each sample.

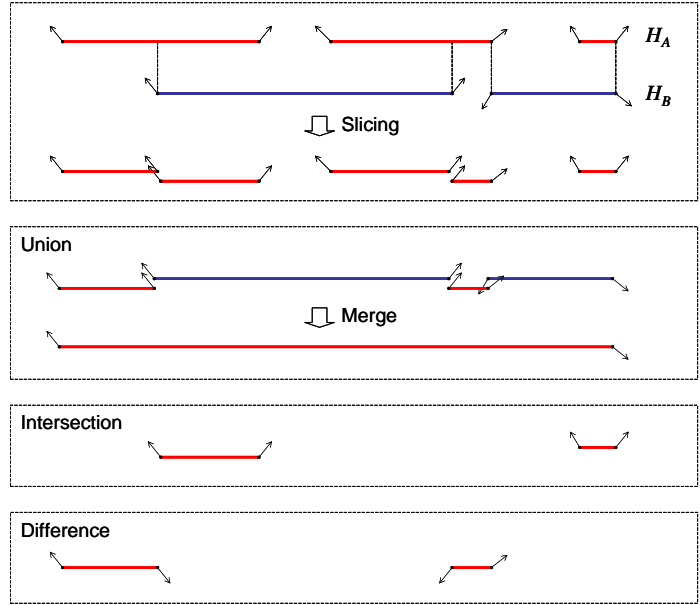

**Figure 5**: Boolean operations on solid models represented by LDNIs are converted into Boolean operations on 1D segments.

## 4. Solid Modeling on LDNI

### 4.1 Boolean operations

As being the fundamental operations of solid modeling, Boolean operations – union, intersection and difference are widely used in various CAD/CAM applications. In LDNI-based representation, a solid model is presented by a set of well-organized 1D volumes (i.e., the even number of depth-normal samples stored in each pixel of LDNIs). When computing the Boolean operation of two solid models $H_A$ and $H_B$, the operations are conducted on the depth-normal samples stored in the corresponding pixels of LDNIs for $H_A$ and $H_B$. The segments on $H_A$ for a pixel are first sliced into more segments by the samples from $H_B$ which fall in the interval of segments on $H_A$ (see the example slicing step shown in Figure 5). Then, we generate the results of Boolean operations as follows.

- *Union*: the segments on $H_A$ that do not overlap the segments on $H_B$ are selected, and merged with the segments on $H_B$ to generate resultant segments.
- *Intersection*: only the segments on $H_A$ that overlap the segments on $H_B$ are remained;
- *Difference*: the segments on $H_A$ that overlap the segments on $H_B$ are removed, and the endpoint sample generated by $H_B$ should have its normal vector inversed.

Illustrations for the operations are given in Figure 5.

#### 4.1.1 Robustness enhancement

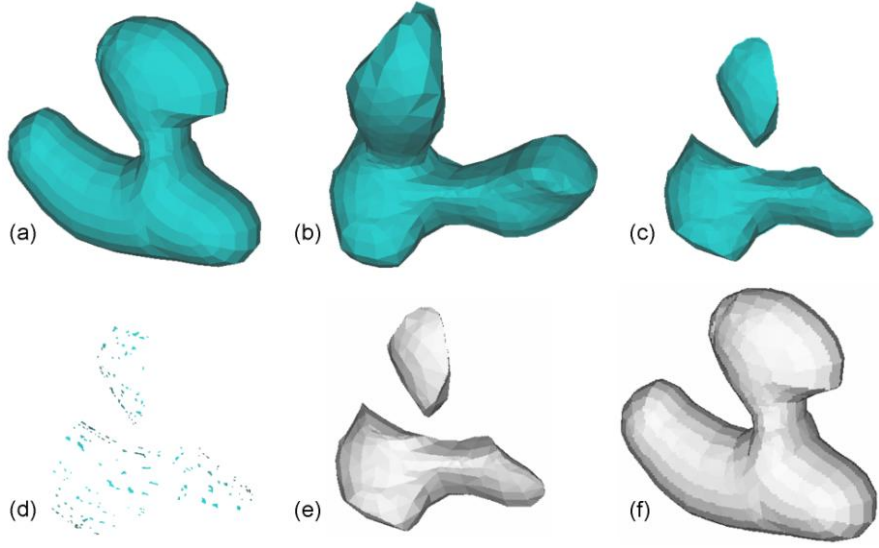

**Figure 6**: Small volume noisy region will be generated on the result of repeating Boolean operations if we keep contouring the model back into B-rep after each operation – (a) the given solid model $H_A$, (b) the given solid model $H_B$, (c) the result of $H_A \cap H_B$ in B-rep, and (d) the result of $(H_A \cap H_B - H_A)$ in B-rep. The robustness of these operations can be improved by keeping a LNDI-solid $L$ and sampling the later coming solid using the same origin and sampling rate of $L$ – (e) $H_A \cap H_B$ in LDNI, (f) the sampling results of $H_A$ in LDNI. The result of $(H_A \cap H_B - H_A)$ in LDNI is then empty, which is consistent with the theoretical result.

A post-processing step – *small interval removal* is given to enhance the robustness of Boolean operation computation. The 1D volumes whose thickness are less than $\varepsilon$ will be removed from the 1D volume of resultant LDNIs (e.g., $\varepsilon = 10^{-5}$ is chosen in our implementation the depth values are encoded in single precision float). By this, the result of Boolean operations on tangential contact models will be corrected.

If repeatedly applying Boolean operations on two solid models and the result is contoured into mesh surface after each Boolean operation (by the contouring method introduced in the next section), many separated region with very small volumes will be produced (as shown in Figure 6(d)). This is because the shape approximation error is generated and accumulated during the boundary surface contouring from models represented by LDNIs. However, such errors can be eliminated if

- All the models in the sequence of Boolean operations are sampled with the same origin and the same sampling rate;
- The temporary models are represented by LDNIs instead of B-rep.

Figure 6(e)-(f) show the results generated by this strategy.

### 4.1.2 Possible GPU based implementation of Boolean operations on LDNIs

As the Boolean operations of solid models in $\Re^3$ have been decomposed into the segments splitting and merging in $\Re^1$, the computation can be easily implemented on GPU which is now considered as a fast and parallel computing power. Also, as will be seen later, the required memory for storing LDNIs in an acceptable resolution is much smaller than the current memory of graphics card. The sampling of LDNIs from polygonal meshes can be easily performed on the GPU side, and the sampling result does not need to be transformed back to the main memory before contouring. This will be a further speed up of the modeling framework introduced in this paper although the current implementation on CPU and main memory has already been able to achieve an interactive speed.

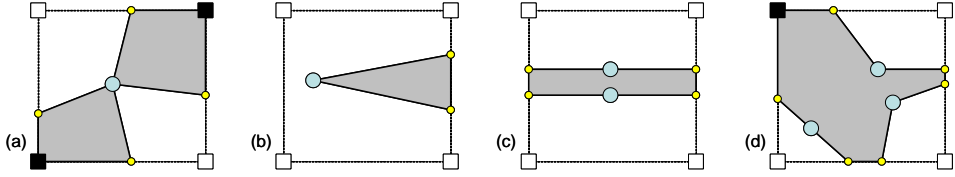

**Figure 7**: The example cases that the original dual contouring algorithm will have incorrect output: (a) a non-manifold mesh surface will be generated, and (b)-(d) the thin-shell will be missed.

### 4.2 Offsetting

Offsetting is another very important operation in CAD/CAM applications. Similar to [9], here the offsetting of a given model *H* with distance *r* is formulated as the Minkowski sum of *H* with a sphere with radius *r*. Computing accurate Minkowski sum of a freeform model and a sphere is impractical and unnecessary. Here, the approximation of Minkowski sum is evaluated. For every depth-normal sample *p* in the LDNIs representation of the given model *H*, we place a sphere *S* centered at *p* where *S* with radius *r* is also in the representation of LDNIs. The union of *H* and *S* is then computed. Repeating the union operations until all depth-normal samples have been processed, we obtain a new model *H\** represented by updated LDNIs. The surface of *H\** approximates the offset surface of *H* with the offset *r* can then be contoured by the algorithm presented below. Similarly, the inner offsetting can be computed by repeatedly subtracting *S* from *H* (i.e., the Boolean difference). Example results are shown in Figure 14.

### 5. Contouring Algorithm: from LDNIs to 2-Manifold Polygonal Mesh

This section will give the contouring algorithm to convert a given solid model *H* from its LDNI-based representation into two-manifold polygonal meshes (i.e., B-rep). The dual contouring algorithm in [3] will generate non-manifold vertices for all the ambiguous sign configurations (ref. [5]) in the original MC algorithm (e.g., the case in Figure 7(a)). Furthermore, the cell node based algorithm will miss the thin structure whose thickness is less than the width of cells (e.g., the cases shown in Figure 7(b)-(d)).

### 5.1. Contouring algorithm with four steps

#### Step 1: Grid node construction

The grid nodes are the intersections of the rays defined by three LDNIs. For a grid node which is an intersection of 3 rays on x-LDNI, y-LDNI and z-LDNI respectively, its *inside/outside* status defined by the three LDNIs is expected to be consistent. However, the numerical error and some degenerated case can lead to inconsistent classification of grid signs. To fix such inconsistency, a grid node is classified into *inside* the model if it falls into the volume defined by at least two of the LDNIs.

**Step 2: Cell edge construction and regularization**

Cell edges are constructed between the grid nodes by the intersections (depth samples) on its corresponding ray of LDNI. Different types of edges are constructed according to the number of depth samples falling in the interval of the edge. The cell edge with its two end-nodes having different sign and with intersection points on the edge is named as *intersect-edge* (e.g., the edges in Figure 7(a)), and the one having different sign nodes but with no intersection point is named as *none-intersect-edge*. The *none-intersect-edge* is generated because of the numerical error. The cell edge with end-nodes having the same sign and no intersection point is an *empty-edge* (e.g., two horizontal edges in Figure 7(c)), and the one with same sign nodes but with intersection point is denoted by *complex-edge* (e.g., two vertical edges in Figure 7(c)). Based on the heuristics that there is only one thin structure passing a cell edge, the following rules are employed to regularize the cell edges.

**Rule 1** A *complex-edge* with more than two intersections will only keep two intersections that are the closest to the end-nodes of the edge.

**Rule 2** An *intersect-edge* with more than one intersection keeps only one sample that is the closest to the middle point of the edge and with the normal vector compatible to the signs on the end-nodes (i.e., the sample with normal's direction consistent with signs on the end-nodes of the edge).

**Step 3: Cell construction and vertices positioning**

A color flooding algorithm is employed to cluster the grid nodes in one cell whose signs indicate *outside*. In each cell, a grid node with *outside* flag is chosen as a seed to fill with a color $c$, and a flooding algorithm is used to fill the color $c$ to all grid nodes in this cell that are linked to this seed by *empty-edges*. If a new *outside* seed node is found on the cell, the flooding is applied again with a new color. This flooding will be repeated until no new seed node is found. The number of vertices to be constructed in this cell depends on the number of colors. After grid nodes with *outside* flag are divided into clusters by different colors, the position of every node cluster is determined by the Hermite data points on the *intersect-edge* and the *complex-edge* linking to this set of grid nodes. The computation is through minimizing the Quadratic Error Function (QEF) defined by these Hermite data points (ref. [3]). For the *complex-edges* with two Hermite data points, each of the Hermite points is classified to the cluster holding one of the two end-nodes. Few vertices associated with no Hermite data point may be generated on the *none-intersect-edges* because of the numerical error. For them, the final positions are determined through Laplacian smoothing after constructing the mesh connectivity.

**Step 4: Mesh construction**

For every regularized *intersect-edge*, one quadrilateral face is constructed by linking the vertices in its four neighboring cells. The adopted vertices should be the one associated with the *outside* node on this intersect-edge. For every regularized *complex-edge*, two quadrilateral faces, each for one end-node on the edge, are constructed in the similar way. All faces should be constructed in the orientation to let its normal facing outwards.

**5.2. Two-manifold correction techniques**

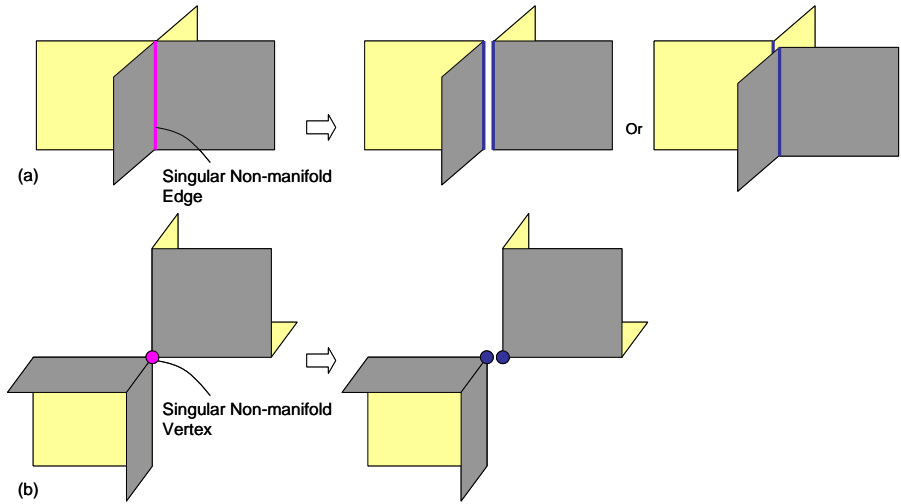

**Figure 8**: Two cases of non-manifold entities generated from the contouring algorithm are corrected. (a) Singular non-manifold edge, where the correction has two options – we choose the right one which is more consistent with our contouring algorithm to have fewer gaps on resultant surfaces. (b) Singular non-manifold vertex and its correction.

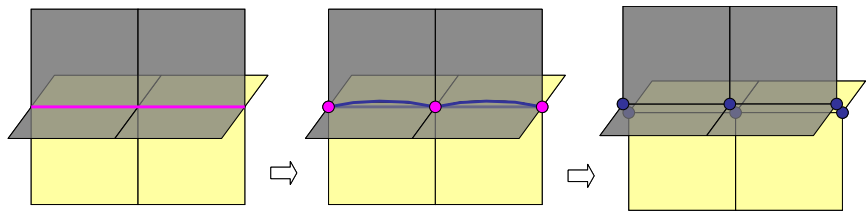

**Figure 9**: An example of progressive two-manifold correction: left – a non-manifold mesh model, middle – singular non-manifold edges are corrected but generate some singular vertices, and right – singular vertices have been corrected.

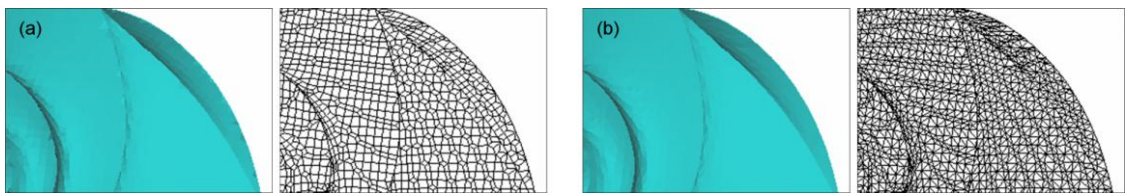

**Figure 10**: Triangulation the quadrilateral faces obtained from the contouring algorithm will give better shape along the sharp edges: (a) before and (b) after triangulation – face shading is adopted.

Similar to the algorithm presented in [9], non-manifold entities will be generated by the above contouring algorithm, which needs to be fixed. Here, the non-manifold entities only have two possible cases (see Figure 8): 1) more than two faces sharing the same polygonal edge – a singular non-manifold edge is found and 2) a vertex is shared by faces that are not connected by manifold edges – a singular vertex is found.

The correction is given in two steps, where all singular non-manifold edges are firstly fixed, and then the faces around singular non-manifold vertices are clustered separately.

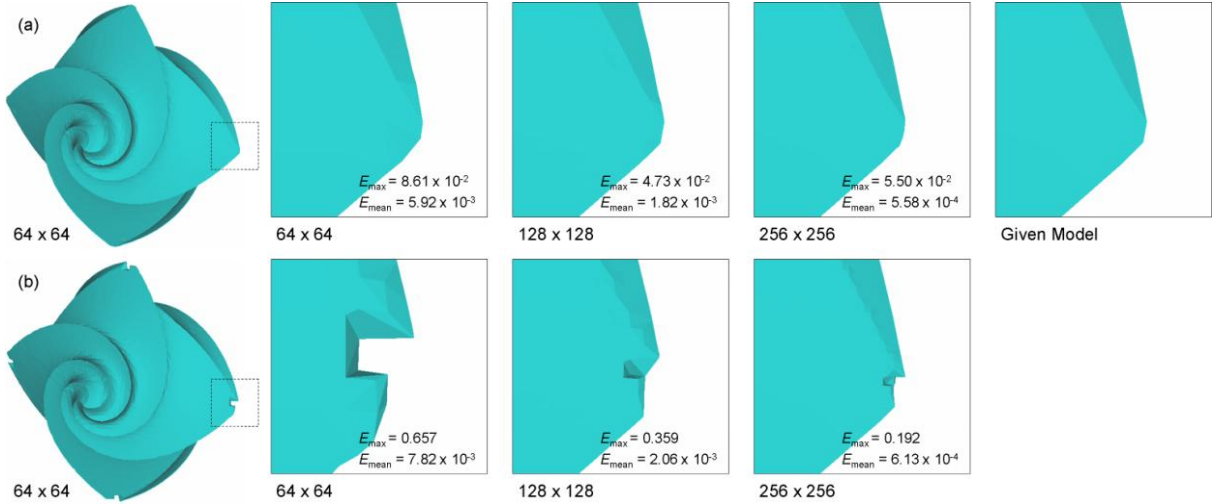

**Figure 11**: The test for comparing the resultant mesh surface generated from LDNIs in different resolutions by (a) the cell-edge based contouring algorithm and (b) the grid node based contouring algorithm [11]. The shape approximation errors (in terms of $E_{max}$ and $E_{mean}$) are measured by the public available Metro tool [44].

- When fixing singular edges, there are two options as shown in Figure 8(a). Our strategy is to let two faces which will form a convex edge being connected.
- When correcting singular vertices, the faces are clustered into groups where faces are linked by manifold edges and then separated by duplicating new vertices (see Figure 8(b)).

The reason why we correct singular edge before correct vertices is because that new singular vertices may be created during the correction of singular edges (e.g., the example shown in Figure 9).

### 5.3. Triangulation of quadrilateral faces

Faces in the resultant surface from the contouring algorithm are quadrilateral. Quadrilateral face may not give a good shape description along sharp features because that the four vertices of a face in general are not coplanar there (e.g., Figure 10(a)). The faces need to be triangulated. During the triangulation, we add a new edge to split a quadrilateral face into two triangles to let the normal variation between the new triangles and their neighbor faces minimized. One example result is given in Figure 10(b), where the surface shape along sharp features has been improved.

### 6. Experimental Results and Discussions

The first test given in this section is to illustrate the difference on resultant mesh surfaces generated from the cell edge based contouring in this paper and the grid node based contouring algorithm as [11]. The surface errors between the contoured mesh surface and the original model are measured by the publicly available Metro tool [44]. Both the maximum surface distance $E_{max}$ and the mean surface distances $E_{mean}$ are reported. From Figure 11, it is easy to find that the thin structures are damaged in the grid node based contouring result since the *complex-edges* are neglected. Although the damaged thin structure can be somewhat compensated with the increasing of sampling rate, the sharp edges still however cannot be fully recovered. Therefore, the cell-edge based contouring algorithm is an important tool to support the LDNI-based implicit representation of solid models.

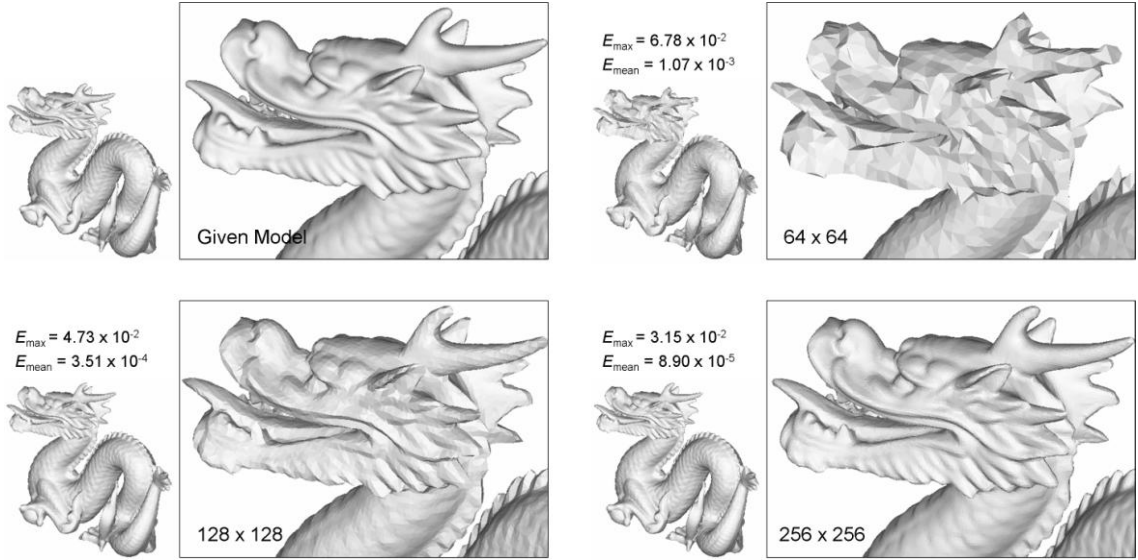

**Figure 12**: The reconstruction result of the dragon model from LDNIs: the given mesh model with 723,708 triangles, the result using $64\times 64$ LDNIs with 16,828 triangles, the result using $128\times 128$ LDNIs with 68,736 triangles, and the result using $256\times 256$ LDNIs with 276,542 triangles.

**Table 1  Analysis of Results from the Accurate Normal vs. the Approximate Normal Sampling**

| Octa-flower | Approximate Normals | | | Accurate Normals | | |
|---|---|---|---|---|---|---|
| | $E_{max}$ | $E_{mean}$ | RAM (MB) | $E_{max}$ | $E_{mean}$ | RAM (MB) |
| Res: $64\times 64$ | $8.61\times 10^{-2}$ | $5.92\times 10^{-3}$ | 0.105 | $8.63\times 10^{-2}$ | $5.91\times 10^{-3}$ | 0.281 |
| Res: $128\times 128$ | $4.73\times 10^{-2}$ | $1.82\times 10^{-3}$ | 0.421 | $4.81\times 10^{-2}$ | $1.81\times 10^{-3}$ | 1.12 |
| Res: $256\times 256$ | $5.50\times 10^{-2}$ | $5.58\times 10^{-4}$ | 1.69 | $5.51\times 10^{-2}$ | $5.52\times 10^{-4}$ | 4.50 |
| **Dragon** | Approximate Normals | | | Accurate Normals | | |
| | $E_{max}$ | $E_{mean}$ | RAM (MB) | $E_{max}$ | $E_{mean}$ | RAM (MB) |
| Res: $64\times 64$ | $6.78\times 10^{-2}$ | $1.07\times 10^{-3}$ | 0.104 | $6.79\times 10^{-2}$ | $1.07\times 10^{-3}$ | 0.277 |
| Res: $128\times 128$ | $4.74\times 10^{-2}$ | $3.51\times 10^{-4}$ | 0.420 | $4.74\times 10^{-2}$ | $3.51\times 10^{-4}$ | 1.12 |
| Res: $256\times 256$ | $3.15\times 10^{-2}$ | $8.90\times 10^{-5}$ | 1.68 | $3.15\times 10^{-2}$ | $8.90\times 10^{-5}$ | 4.46 |

Our second test is to investigate whether an accurate normal sampling is important. As shown in Table 1, two models – the octa-flower and the dragon are sampled into LDNIs with approximate normal vectors (directly sampled from RBG colors and stored in 3 bytes for each vector) versus accurate normal vectors (recorded from the original polygonal faces through the color-based index querying and stored in 3 double precision variables for each vector). Then, the mesh surfaces of the models are reconstructed by our sharp-feature preserved contouring algorithm and measured comparing to the input model by the Metro tool [44]. The test is conducted in three different resolutions. From the error analysis in Table 1, we can conclude that there is not too much difference between the results from the accurate and the approximate normal sampling in terms of surface error.

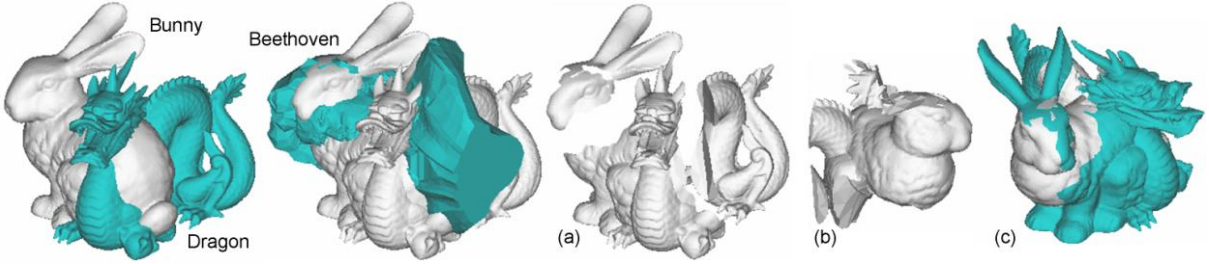

**Figure 13**: Results of Boolean operations: (a) (Bunny ∪ Dragon) – Beethoven, (b) (Bunny ∪ Dragon) ∩ Beethoven, and (c) placing (a) & (b) together.

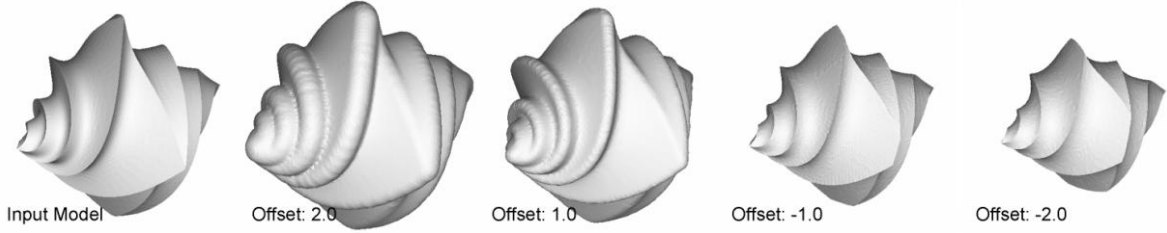

**Figure 14**: The example results of offsetting: the computation can be finished in 30 – 70 seconds by using LDNIs with the resolution: 128×128.

However, the memory required for storing the LDNIs with accurate normals is more than double of the memory occupied by LDNIs with approximate normal vectors. Therefore, LDNIs with approximate normals are used for all the other examples shown in this paper. Figure 11(a) and 12 show the polygonal meshes for the octa-flower and the dragon model contoured from LDNIs with approximate normal vectors.

We have implemented the proposed modeling framework by both 1) GLUT and 2) the MFC based off-screen rendering on DIB image, where the previous implementation take advantage of the graphics hardware acceleration and the late one fully depends on the software implementation. The time differences of remeshing on these two implementations have been listed in Table 2. It is clear that the GLUT based implementation is much faster – in some cases, it speed up more than 2/3 of the sampling time of LDNIs because of the graphics hardware. Actually, the time could be further shortened by the newly provided functions on modern graphics card (see the discussion in section 6.2).

Figure 13 shows the results of Boolean operations given on three freeform solid models: the Bunny, the Dragon and the Beethoven. The Boolean operations implemented on LDNIs are very robust and stable. When placing the resultant models in Figures 13(a) and 13(b) together, it is almost seamless. Statistics of the computation are given in Table 3. The computations are tested on LDNIs in two different resolutions – 128×128 and 256×256. From the statistic, it is easy to find that the most time-consuming step is the contouring as our contouring algorithm is in the cubic complexity as the sampling density $w$. Note that, if repeated Boolean operations will be applied, we actually only need to contour the boundary surface of resultant solid models after completing all Boolean operations.

**Table 2  Remeshing Time of the Implementations on GLUT vs. MFC Off-screen Rendering**

| GLUT rendering based | | | | | | |
|---|---|---|---|---|---|---|
| **Models** | Time of Res.: 64×64 | | Time of Res.: 128×128 | | Time of Res.: 256×256 | |
| | Sampling | Contouring | Sampling | Contouring | Sampling | Contouring |
| Octa-flower (8) | 0.047 s | 0.485 s | 0.125 s | 2.875 s | 0.234 s | 20.846 s |
| Dragon (14) | 5.797 s | 1.328 s | 7.916 s | 7.906 s | 9.725 s | 24.372 s |
| Gear (10) | 0.016 s | 0.671 s | 0.093 s | 3.563 s | 0.219 s | 27.294 s |
| **MFC off-screen rendering based** | | | | | | |
| **Models** | Time of Res.: 64×64 | | Time of Res.: 128×128 | | Time of Res.: 256×256 | |
| | Sampling | Contouring | Sampling | Contouring | Sampling | Contouring |
| Octa-flower (8) | 0.281 s | 0.922 s | 0.391 s | 4.093 s | 0.734 s | 27.704 s |
| Dragon (14) | 23.359 s | 1.891 s | 21.407 s | 5.328 s | 29.156 s | 31.391 s |
| Gear (10) | 0.203 s | 1.688 s | 0.281 s | 5.922 s | 0.609 s | 38.141 s |

\* The computations are given on a standard PC with Pentium IV 3.0 GHz CPU + 2GB RAM and a NVIDIA GeForce 7600 GS graphics card with 256MB RAM running Windows XP; the number in bracket is the maximal number of layers in LDNIs. The numbers of faces on the models are the Octa-flower (15,843), the Dragon (723,708) and the Gear (6,626).

**Table 3  Computational Statistic of LDNI-based Boolean Operations**

| LDNIs in resolution: 128×128 | | | | | | | |
|---|---|---|---|---|---|---|---|
| Model | Input Trgl # | Boolean Type | Time of | | | LDNI Memory | Output Trgl # |
| | | | LDNI Sampling | Boolean Operation | Contouring | | |
| Bunny | 69,664 | Union | 9.421 s | 0.047 s | 5.95 s | 0.84MB | 98,984 |
| Dragon | 723,708 | | | | | | |
| Union | 98,984 | Difference | 0.973 s | 0.046 s | 3.31 s | 0.93MB | 91,374 |
| Beethoven | 5,030 | | | | | | |
| Union | 98,984 | Intersection | 0.973 s | 0.048 s | 2.39 s | 0.79MB | 50,560 |
| Beethoven | 5,030 | | | | | | |
| **LDNIs in resolution: 256×256** | | | | | | | |
| Model | Input Trgl # | Boolean Type | Time (in sec) of | | | LDNI Memory | Output Trgl # |
| | | | LDNI Sampling | Boolean Operation | Contouring | | |
| Bunny | 69,664 | Union | 13.638 s | 0.308 s | 19.94 s | 3.36MB | 399,238 |
| Dragon | 723,708 | | | | | | |
| Union | 399,238 | Difference | 7.916 s | 0.251 s | 30.38 s | 3.75MB | 372,452 |
| Beethoven | 5,030 | | | | | | |
| Union | 399,238 | Intersection | 7.916 s | 0.203 s | 25.35 s | 3.20MB | 206,464 |
| Beethoven | 5,030 | | | | | | |

\*The computations are given on a standard PC with Pentium IV 3.0 GHz CPU + 2GB RAM and a NVIDIA GeForce 7600 GS graphics card with 256MB RAM

**Table 4  Computational Statistic of Offsetting**

| Flower Model | Triangle Number | Sampling Time | Offsetting Time | Contouring Time |
|---|---|---|---|---|
| **Offset** $r = 2.0$ | 36,420 | 0.235 s | 66.25 s | 2.58 s |
| **Offset** $r = 1.0$ | 37,314 | 0.234 s | 28.53 s | 2.64 s |
| **Offset** $r = -1.0$ | 35,498 | 0.250 s | 38.97 s | 2.79 s |
| **Offset** $r = -2.0$ | 15,078 | 0.219 s | 34.17 s | 1.72 s |

\*The computations are given on the same PC as above, and the input model has 15,834 triangles.

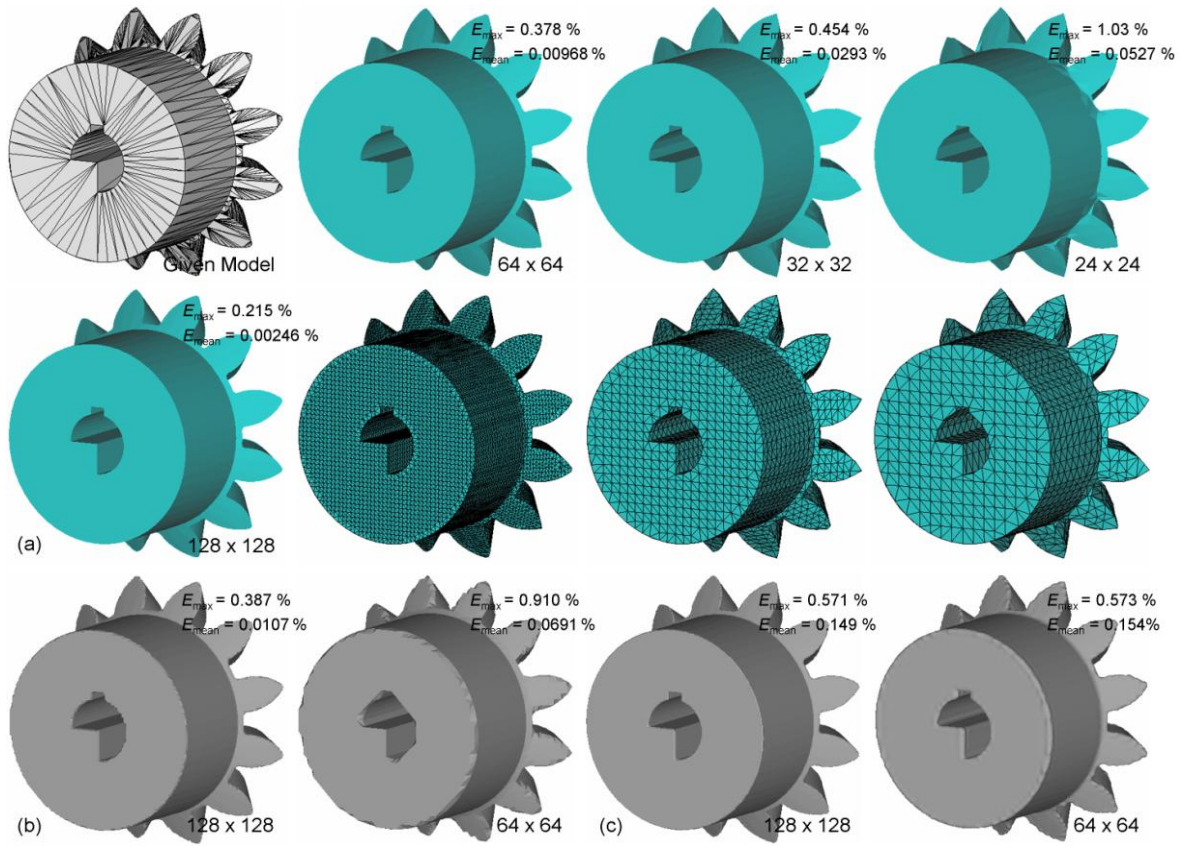

**Figure 15**: Reconstruction of gear model using different implicit representations: (a) our method based on LDNIs presented in this paper with different resolutions, (b) the mesh surface generated by Marching Cubes algorithm [8] from uniformly sampled distance field, and (c) the implicit surface generated by the method in [45] where the samples from LDNIs are used as oriented point cloud as input. The shape approximation errors are measured by the publicly available Metro tool [44] with reference to the diagonal length of the original gear model.

The offsetting operation based on computing the approximated Minkowski sum by LDNIs is also very stable and fast. Figure 14 gives the offsetting results of an input flower model with offset 2.0, 1.0, -1.0 and -2.0. As our Boolean operations can be robustly and effectively completed, sharp features are preserved on the offsetting results, which can hardly be produced by some other implicit representation based approaches like Level-set approaches [2]. The computational statistic for offsetting is given in Table 4.

### 6.1 Comparison to other approaches

We have compared the reconstruction result from the LDNI representation proposed in this paper with several other implicit representations in Figure 15. First, the uniform distance field is used to sample the given gear model. The reconstruction results are then generated by the Marching Cubes algorithm [8]. It is easy to found that the sharp edges are chamfered (see Figuure 15(b)). Secondly, the Hermite samples from LDNIs are employed as oriented point cloud to reconstruct an implicit surface for the gear model by the method of Ohtake et al. [45], which cannot recover sharp features too.

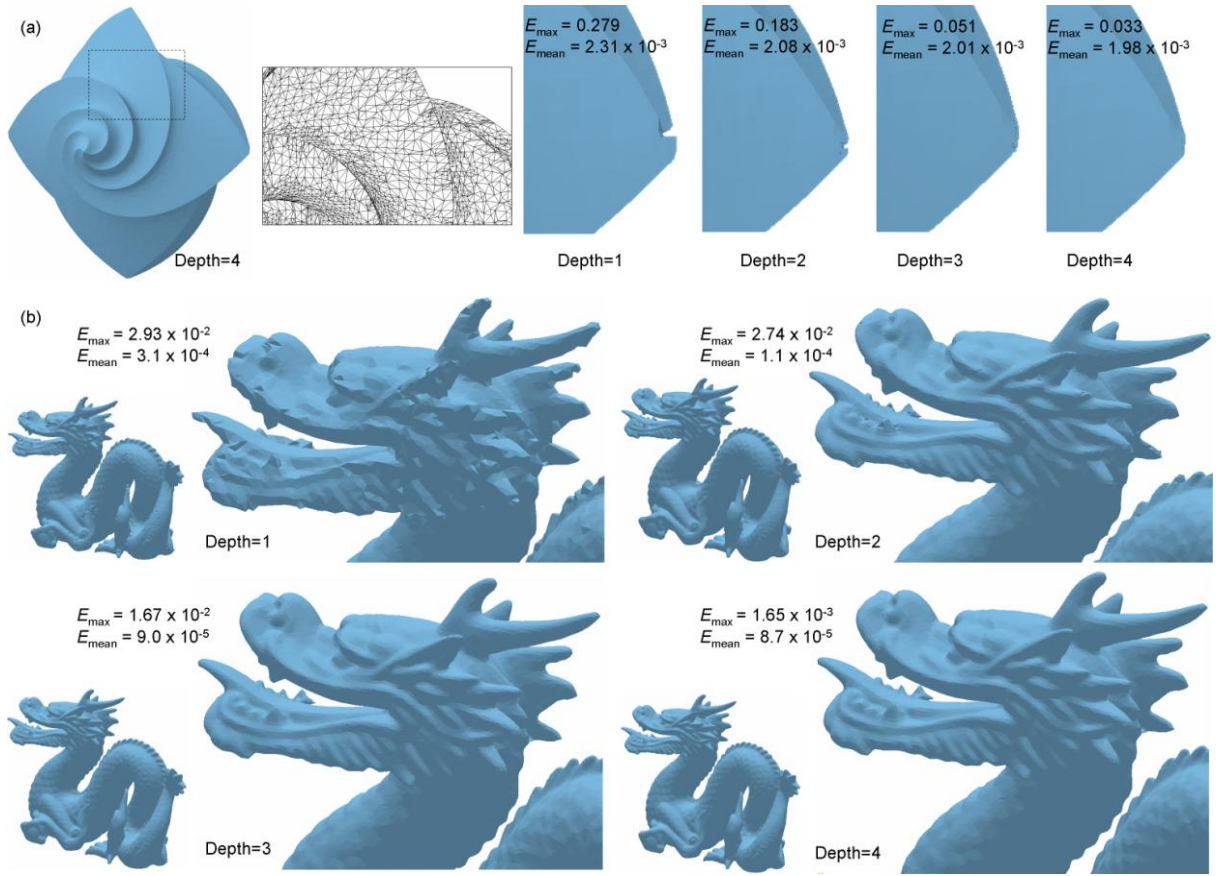

**Figure 16**: The remeshing results of (a) the Octa-flower model and (b) the Dragon model by the adaptive Octree in depth 1-4 using the method in [19].

Our reconstruction results have also been compared with the results from the adaptive sampling based approach [19] on two models – the Octa-flower and the Dragon. By [19], a grid size 0.025 inch is adopted to sample the models, and the generated uniform grids are $56 \times 55 \times 40$ and $50 \times 35 \times 22$ for the Octa-flower and Dragon models respectively. This was followed by Octree grids with a maximum depth of 1, 2, 3 and 4 to refine the given models. The reconstruction results are shown in Figure 16, and the statistics in different resolutions are given in Table 5.

The reason why we choose depth 1-4 is because that when starting from the uniform grids around 50, Octree at depth 1 is with similar sampling rate as $128 \times 128$ LDNIs and Octree at depth 2 is similar to $256 \times 256$ LDNIs. The errors in terms of $E_{max}$ and $E_{mean}$ on remeshed models from Octree-based adaptive sampling at depth 1 and 2 are similar to the results remeshed from $128 \times 128$ and $256 \times 256$ LDNIs respectively. However, comparing the size of required memory under a similar sampling rate, the LDNI approach needs much less memory than [19] (especially on the dragon model which is much more complex). When considering about the sampling time, the LDNI-based approach is obviously faster than the adaptive sampling approach as the graphics hardware acceleration is adopted. Whether the adaptive sampling approach can also borrow the power of graphics hardware is still an open problem to be investigated.

**Table 5  Computational Statistics of Adaptive Sampling Based Remeshing**

| Octa-flower | | | | | | | |
|---|---|---|---|---|---|---|---|
| Octree Depth | Time (in sec) of Constructing | | Contouring Time (sec) | Octree Memory | Output Trgl # | $E_{max}$ | $E_{mean}$ |
| | Uniform_Grids | Octree_Grids | | | | | |
| 1 | 1.3 s | 5.4 s | 0.3 s | 1.8 MB | 31,718 | 0.279 | $2.3 \times 10^{-3}$ |
| 2 | 1.3 s | 6.5 s | 0.4 s | 2.2 MB | 37,886 | 0.183 | $2.1 \times 10^{-3}$ |
| 3 | 1.3 s | 7.0 s | 0.4 s | 2.4 MB | 39,956 | 0.051 | $2.0 \times 10^{-3}$ |
| 4 | 1.3 s | 7.3 s | 0.5 s | 2.5 MB | 41,296 | 0.033 | $2.0 \times 10^{-3}$ |
| **Dragon** | | | | | | | |
| Octree Depth | Time (in sec) of Constructing | | Contouring Time (sec) | Octree Memory | Output Trgl # | $E_{max}$ | $E_{mean}$ |
| | Uniform_Grids | Octree_Grids | | | | | |
| 1 | 486 s | 91 s | 0.5 s | 16 MB | 47,024 | $2.93 \times 10^{-2}$ | $3.1 \times 10^{-4}$ |
| 2 | 488 s | 144 s | 1.3 s | 27 MB | 105,526 | $2.74 \times 10^{-2}$ | $1.1 \times 10^{-4}$ |
| 3 | 487 s | 162 s | 1.5 s | 30 MB | 125,760 | $1.67 \times 10^{-2}$ | $9.0 \times 10^{-5}$ |
| 4 | 486 s | 164 s | 1.6 s | 30 MB | 130,885 | $1.65 \times 10^{-2}$ | $8.7 \times 10^{-5}$ |

*The computations are given on a standard PC with Pentium IV 3.2 GHz CPU + 2GB RAM.

In summary, LDNI-based approach has the following advantages:

- Less memory is required for storing complex models under the similar approximating error;
- The data structure retains the information of complex-edges so that it can reconstruct sharp-features or thin structures in a relative low sampling rate;
- The data structure and relevant algorithms are well structured so that it can easily employ the power of the accelerated graphics hardware (e.g., GPU) or the parallel computing.

**6.2 Discussion**

The upper bound of memory occupied by LDNIs for presenting a solid model is in the complexity $O(n_{max}w^2)$, where $n_{max}$ is the maximal complexity of layers. In practical applications, the value of $n_{max}$ is in a constant complexity. Therefore, the complexity of LDNIs representation is quadratic on most models. However, for the worst case that $n_{max} \to w$, the complexity becomes cubic.

The minimal cell width that can be represented by LDNIs with the resolution $w \times w$ is $1.0/w$. However, different from other grid-nodes based methods, ours can record and reconstruct the features that are smaller than $1.0/w$ in one direction, such as a thin-shell or sharp-feature passing a cell edge. Our method cannot generate the boundary of a solid model with multiple thin-shells (or sharp-feature) that the distance between which is less than $1.0/w$. In addition, our method lacks of the ability to fully process the features that are smaller than $1.0/w$ in two directions, such as small extrusions or small deep holes (e.g., the model shown in Figure 17). This is caused by the aliasing error during sampling. In comparison, the adaptive sampling based approach will adaptively use a sufficiently small cell size to sample the regions with these features (refer to an example shown in Figure 17).

The aliasing from sampling will also bring a limitation on the LDNI-based offsetting operation. When the offset $r$ is not more than $1.0/w$, the quality of result surface will be worse. This is because that, the sample of the sphere with radius $r$ is just one pixel in each LDNI when $r \leq 1.0/w$, which will generate many burrs on the resultant surface (see Figure 18). With the increasing of $r$, the situation will be improved.

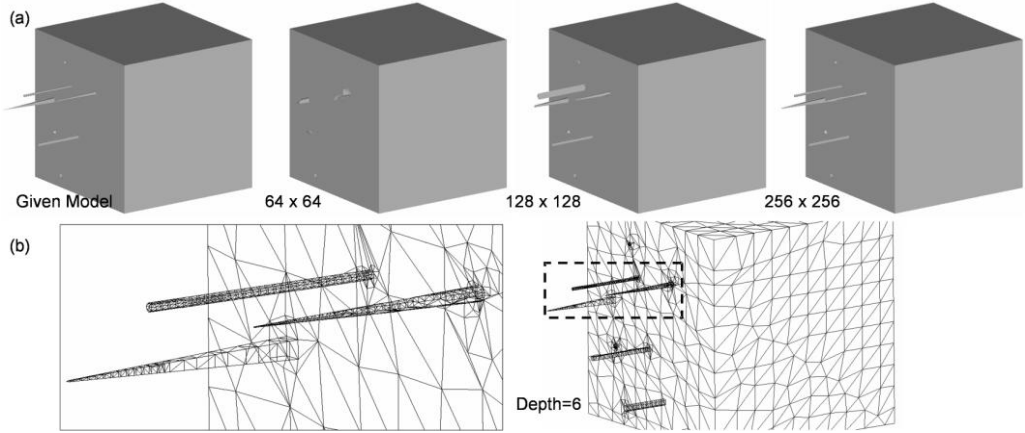

**Figure 17**: The features that are smaller than $1.0/w$ in two directions will be damaged in LDNIs based implicit representation. (a) the remeshing results with LDNIs in res.: $64\times64$ and $128\times128$ have the small features degraded. (b) the adaptive sampling based approach will use a smaller cell size to sample the regions with these features – the remeshing results generated by [19] for a maximum depth of 6 is shown.

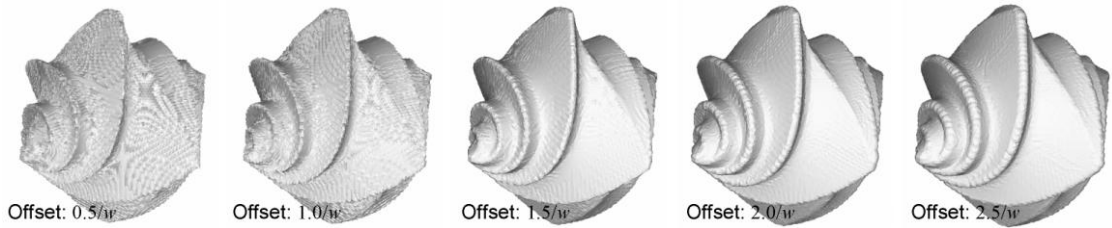

**Figure 18**: The offsetting surface will have bad surface quality with small offset – $r \leq 1.0/w$. While increasing the offset $r$, the situation has been improved.

The contouring algorithm presented in this paper becomes slow when using LDNIs with large resolutions. Therefore, an adaptive contouring algorithm as [46] can be considered to speed up the contouring step and reduce the number of triangles on resultant meshes. Furthermore, the mesh generation in each cell is independent to other cells. This good property will make it easy to develop some parallel algorithm running on GPU or Multi-core CPU.

Our sampling step is implemented with the help of graphics hardware; but the sampled results are read back into the main memory, which actually becomes the bottleneck of our implementation. This is the reason why the sampling is only speed up at about 2/3. Moreover, for the models with many layers (e.g., the Dragon), many pixels for invalid samples are also read back by the code in Appendix A. We can consider about using GPU to eliminate those invalid samples before sending back the Hermite samples of LDNIs, or developing fully GPU based implementations of solid modeling operation. By this, we can further speed up the computation. The sampling, modeling, and contouring steps of LDNIs are all well structured so that they can be further developed into fully GPU based implementations to take advantage of the fast and parallel computing power nowadays provided by GPU. We will consider this in our future research and development.

## 7. Conclusion

A novel sparse implicit representation of solid models, LDNIs, has been introduced in this paper, by which every solid model can be presented by three layered depth-normal images (LDNIs) that are perpendicular to three orthogonal axes respectively. The representation of LDNIs is very effective to capture thin-structures (or sharp-features) whose size is smaller than the distance between two sampling points. The new implicit representation LDNIs has been equipped with the sampling algorithm, the contouring algorithm and the solid modeling operations which are presented in this paper. The computation for these algorithms and operations can be finished in an interactive speed. Comparing to other existing implicit representation of solid models, LDNIs requires less memory. The encoded data of LDNIs for a solid model is well structured so that we can take advantage of the accelerated graphics hardware and the parallel computing power. The solid modeling operations based on LDNIs can be performed efficiently and robustly.

**Acknowledgement**

This work is partially supported by the HKSAR RGC/CERG Research Grant CERG/416307 and CERG/417508 and the CUHK DAG Research Grant CUHK/2050400. The second author would also acknowledge the support of the James H. Zumberge Faculty Research and Innovation Fund at the USC. The authors would also like to thank Dr. Shengjun Liu and Ms. Jiayi Xu for helping generate some figures in the paper.

# Appendix A  Pseudo-Code for the implementation of LDNI sampling in OpenGL

The implementation of LDNI sampling in one direction akin to [24] is as follows.

```
// Rendering setup steps;
Compiling a glList for the model to be sample;
glEnable(GL_DEPTH_TEST);
glEnable(GL_STENCIL_TEST);
Disable all lighting and anti-aliasing functions;
glViewport(0, 0, res, res);    // res is the resolution of LDNI sampling
glMatrixMode(GL_PROJECTION);
glLoadIdentity();
glOrtho(-ww*0.5, ww*0.5, -ww*0.5, ww*0.5, -ww*0.5, ww*0.5);    // ww is the width of bounding cube
glMatrixMode(GL_MODELVIEW);
glLoadIdentity();

// Pass 1 computes n_max and the first LDNI layer
glClearColor( 1.0f, 1.0f, 1.0f, 1.0f );       glClearDepth(1.0);
glClearStencil(0);
glClear( GL_COLOR_BUFFER_BIT | GL_DEPTH_BUFFER_BIT | GL_STENCIL_BUFFER_BIT);
glDepthFunc(GL_ALWAYS);
glStencilFunc(GL_GREATER, 1, 0xff);
glStencilOp(GL_INCR, GL_INCR, GL_INCR);
Rendering object by calling the pre-complied glList;
Depth complexities ← stencil buffer;
n_max ← max(depth complexities);
Layer [1] ← depth buffer + color buffer;

// Passes 2 to n_max compute the remaining LDNI layers
n ← 2;
While n ≤ n_max Begin
    glClear( GL_COLOR_BUFFER_BIT | GL_DEPTH_BUFFER_BIT | GL_STENCIL_BUFFER_BIT);
    glStencilFunc(GL_GREATER, n, 0xff);
    glStencilOp(GL_ KEEP, GL_INCR, GL_INCR);
    Rendering object by calling the pre-complied glList;
    Layer [n] ← depth buffer + color buffer;    // only those depth complexity ≤ n are recorded
    n ← n + 1;
End
```